\documentclass[aps,preprint,showpacs,nobibnotes]{revtex4-1}
\pdfoutput=1
\usepackage{graphics,graphicx,amsfonts,amsmath,amsbsy,amssymb,color}
\usepackage{verbatim}  
\usepackage{psfrag}  
\usepackage{tabularx}
\usepackage{xmpmulti}
\usepackage{hyperref}
\usepackage{titlesec}
\usepackage{multirow}
\usepackage{subfigure}
\usepackage{marginnote}
\usepackage{enumitem}
\usepackage{epstopdf}
\usepackage{placeins}

\usepackage{hyphenat}
\hypersetup{hyperindex,breaklinks,colorlinks=true,linkcolor=blue,citecolor=blue,urlcolor=blue,filecolor=blue}

\newcommand{\refeq}[1]{{Eq.~(\ref{#1})}}
\newcommand{\reffig}[1]{{Fig.~\ref{#1}}}

\begin{document}

\author{Nadav Geva$^{(1)}$}
\author{James J. Shepherd$^{(2)}$}
\author{Lea Nienhaus$^{(2)}$}
\author{Moungi G. Bawendi$^{(2)}$}
\author{Troy Van Voorhis$^{(2)}$}
\email{tvan@mit.edu}
\affiliation{$^{(1)}$Department of Materials Science and Engineering, Massachusetts Institute of Technology, Cambridge, Massachusetts 02139-4307}
\affiliation{$^{(2)}$Department of Chemistry, Massachusetts Institute of Technology, Cambridge, Massachusetts 02139-4307}

\title{Morphology of passivating organic ligands around a nanocrystal}

\begin{abstract}

Semiconductor nanocrystals are a promising class of materials for a variety of novel optoelectronic devices, since many of their properties, such as the electronic gap and conductivity, can be controlled. Much of this control is achieved via the organic ligand shell, through control of the size of the nanocrystal and the distance to other objects. 
We here simulate ligand-coated CdSe nanocrystals using atomistic molecular dynamics, allowing for the resolution of novel structural details about the ligand shell.
We show that the ligands on the surface can lie flat to form a highly anisotropic `wet hair' layer as opposed to the `spiky ball' appearance typically considered.
We discuss how this can give rise to a dot-to-dot packing distance of one ligand length since the thickness of the ligand shell is reduced to approximately one-half of the ligand length for the system sizes considered here; these distances imply that energy and charge transfer rates between dots and nearby objects will be enhanced due to the thinner than expected ligand shell.
Our model predicts a non-linear scaling of ligand shell thickness as the ligands transition from `spiky' to `wet hair'. We verify this scaling using TEM on a PbS nanoarray, confirming that this theory gives a qualitatively correct picture of the ligand shell thickness of colloidal quantum dots.
\end{abstract}
\maketitle

Semiconducting quantum dots have attracted substantial attention due to their tunable structure-property relationships~\cite{kagan_electronic_1996,puzder_effect_2004-1,zherebetskyy_hydroxylation_2014,peng_shape_2000,choi_exploiting_2016, moreels_size-dependent_2009}. 
The ability to simultaneously engineer their electronic and optical properties within a single device has made them a prime candidate material in a variety of applications.
In solar cells and LEDs, quantum dot size is used to tune band gaps, and this is commonly exploited to produce varied spectral properties~\cite{kramer_architecture_2014, pattantyus-abraham_depleted-heterojunction_2010,caruge_colloidal_2008, seo_fully_2014,curutchet_examining_2008}. 

These optoelectronic devices function through electronic processes that are often strongly dependent on distance~\cite{collier_reversible_1997,remacle_gating_2003, shirasaki_emergence_2013, guzelturk_organicinorganic_2015, choi_exploiting_2016}. 
For example, conductivity in a quantum dot array is mediated by Marcus-type charge transfer events between dots\cite{marcus_theory_1956, yu_n-type_2003, remacle_gating_2003, law_structural_2008, caruge_colloidal_2008, liu_dependence_2010, weidman_interparticle_2015, akselrod_subdiffusive_2014, kamat_quantum_2008}.
As the dot-to-dot distance increases, the charge transfer rate decays exponentially, making the conductivity extremely sensitive to the dot-to-dot distance~\cite{marcus_theory_1956, chandler_electron_2007, zabet-khosousi_charge_2008, yu_n-type_2003}.
Excitonic energy transfer, relevant in solar cells and light emitters, usually occurs through Forster resonant energy transfer (FRET)\cite{forster_zwischenmolekulare_1948, prins_reduced_2014, mork_magnitude_2014, akselrod_subdiffusive_2014, lee_determination_2015, gil_excitation_2016, kagan_long-range_1996, kagan_electronic_1996} or Dexter processes~\cite{dexter_theory_1953, mongin_direct_2016, thompson_energy_2014}; these are also dependent on distance.
Since the organic ligand shell is usually composed of insulating alkane chains, they behave as a spacer layer that can determine that closest approach distance~\cite{remacle_gating_2003, akselrod_subdiffusive_2014, weidman_interparticle_2015, huang_surface_2005, liu_dependence_2010}. 
Ligand exchange reactions~\cite{anderson_ligand_2013, dubois_versatile_2007, zarghami_p-type_2010, brown_energy_2014} give us \emph{in situ} synthetic access to the ligand shell, and using this design space it is possible to achieve fine control of the aforementioned electronic processes.

Recently, the ability to control the energy gap and energy transfer has been exploited for novel optoelectronic devices~\cite{pattantyus-abraham_depleted-heterojunction_2010,caruge_colloidal_2008,seo_fully_2014}, allowing for down-conversion of a high energy UV photon into two lower energy photons~\cite{thompson_energy_2014, prins_reduced_2014, mongin_direct_2016}, and up-conversion of two low energy IR photons into one higher energy photon~\cite{wu_solid-state_2016, mahboub_triplet_2016, tabachnyk_resonant_2014-2}. 
The ability to up-, and down-convert photon energy can allow solar cells to capture more of the solar spectrum, thereby circumventing the Shockley-Queisser limit~\cite{shockley_detailed_1961,ellingson_highly_2005, lunt_practical_2011}.
However, these conversion processes rely on the aforementioned energy transfer mechanisms and, therefore, are very sensitive to the structure and thickness of the ligand shell\cite{hetsch_semiconductor_2011, rogach_energy_2009}.

In view of the importance of distance-based phenomena, we here address the physical structure of the ligand shell. 
The morphology of the ligand shell is hard to access in experiments and is usually inferred through measurements of the dot-to-dot geometries~\cite{weidman_interparticle_2015, sun_bright_2012, curutchet_examining_2008}.
Detailed atomistic simulations that account for all the ligands and the atoms in the quantum dot can give us direct information on the morphology of the organic ligand shell\cite{blonski_molecular_1993, nakano_dynamics_1995, ramalingam_evolution_2001}. Some previous work ~\cite{djebaili_atomistic_2015,djebaili_atomistic_2013, yadav_effective_2016, heikkila_atomistic_2012, bolintineanu_effects_2014} has looked at tethered ligands on metallic nanocrystals but have not focused specifically on the ramifications of their results on the ligand morphology. 
Without that understanding, it is difficult to make quantitative predictions about the rates of distance dependent processes in quantum dots.

In this work, we present a study where we examine the structure of the ligand shell using a combination of computational simulations, electron microscopy, and theoretical analysis. 
In particular, we undertook a molecular dynamics study on the amine ligand shell surrounding a CdSe nanocrystal, where both the core and the ligand shell are treated atomistically. %
In so doing, we found that the ligands show a tendency to lay flat against the surface, leading to an effective shell thickness reduction over the range of ligand lengths studied (0.3nm~to 2.5nm). 
Quantum dots of nanocrystalline PbS were then synthesized with carboxylic acid ligands and the dot-to-dot distance was measured by transmission electron microscopy.
We were able to use these data to verify our simulations for ligand shell thickness, whilst also demonstrating the transferability of our theoretical findings.
Theoretical analysis was then used to understand the physical chemistry of the ligands in terms of the microscopic properties of the ligands, through the dihedral angle. 
Our results imply that energy and electron transfer processes involving dots may be significantly enhanced in practice due to the unique morphology of the ligand shell.

\section*{Simulations} 

We simulate the morphology of the ligand shell by performing classical molecular dynamics with atomistic detail.
We chose CdSe quantum dots with amine ligands due to the quality of the experimental data~~\cite{munro_quantitative_2007, anderson_ligand_2013, ekimov_absorption_1993, valdez_low_2014}, for the ease of simulation (\emph{i.e.} amines are not charged) and because force fields are readily available\cite{jorgensen_development_1996, schapotschnikow_adsorption_2009}. 
The amine head group can only have one attachment site, as opposed to the two attachment sites in oleic acid ligands, which facilitates ligand placement.
The amines we study have a simple alkane chain  backbone of between three and 19 carbon atoms (CH$_3$(CH$_2$)$_n$NH$_2$).

\begin{figure}

\makebox[0.5\textwidth][l]{%
\subfigure[\mbox{}]{%
\includegraphics[width=0.25\textwidth]{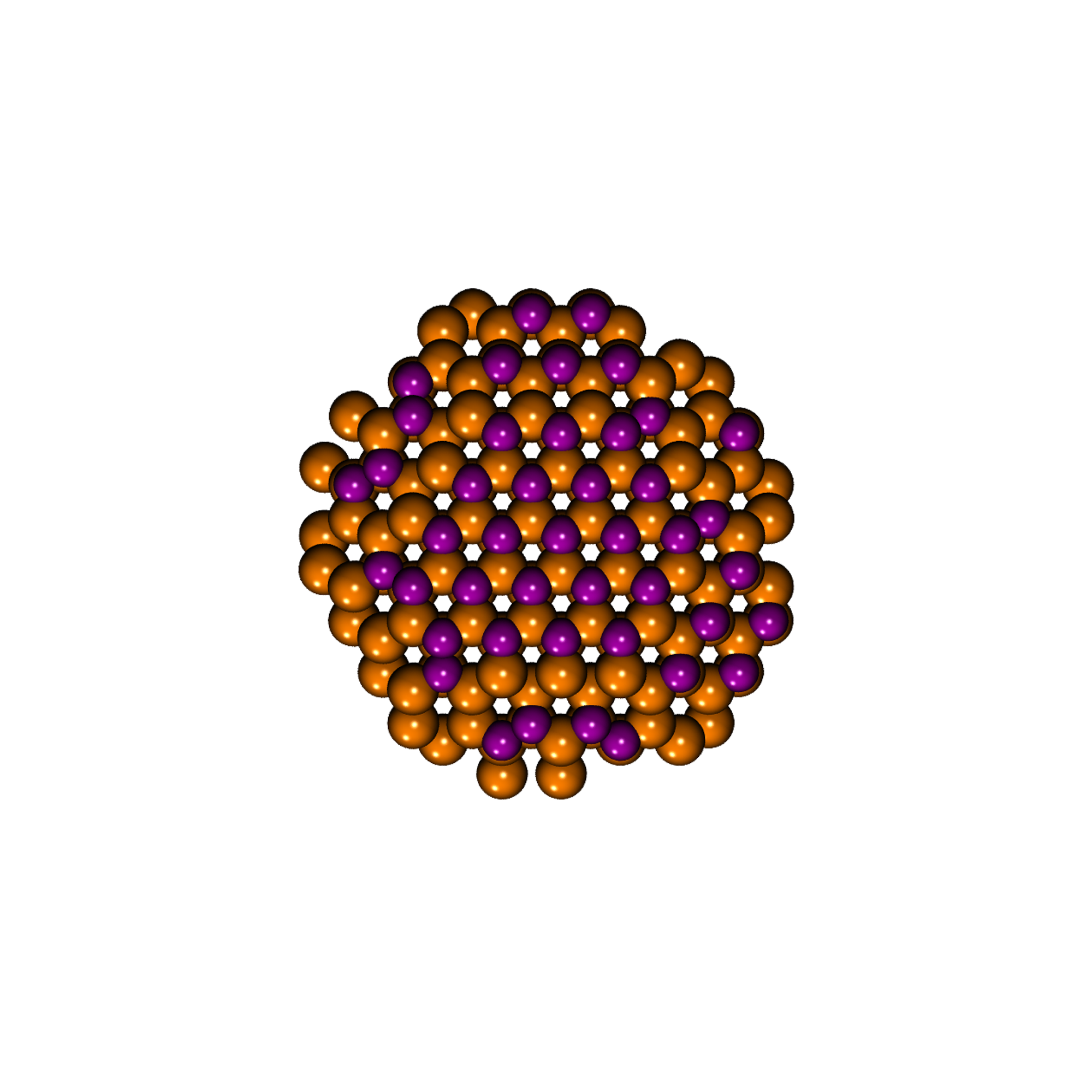}%
}%
\subfigure[\mbox{}]{%
\includegraphics[width=0.25\textwidth]{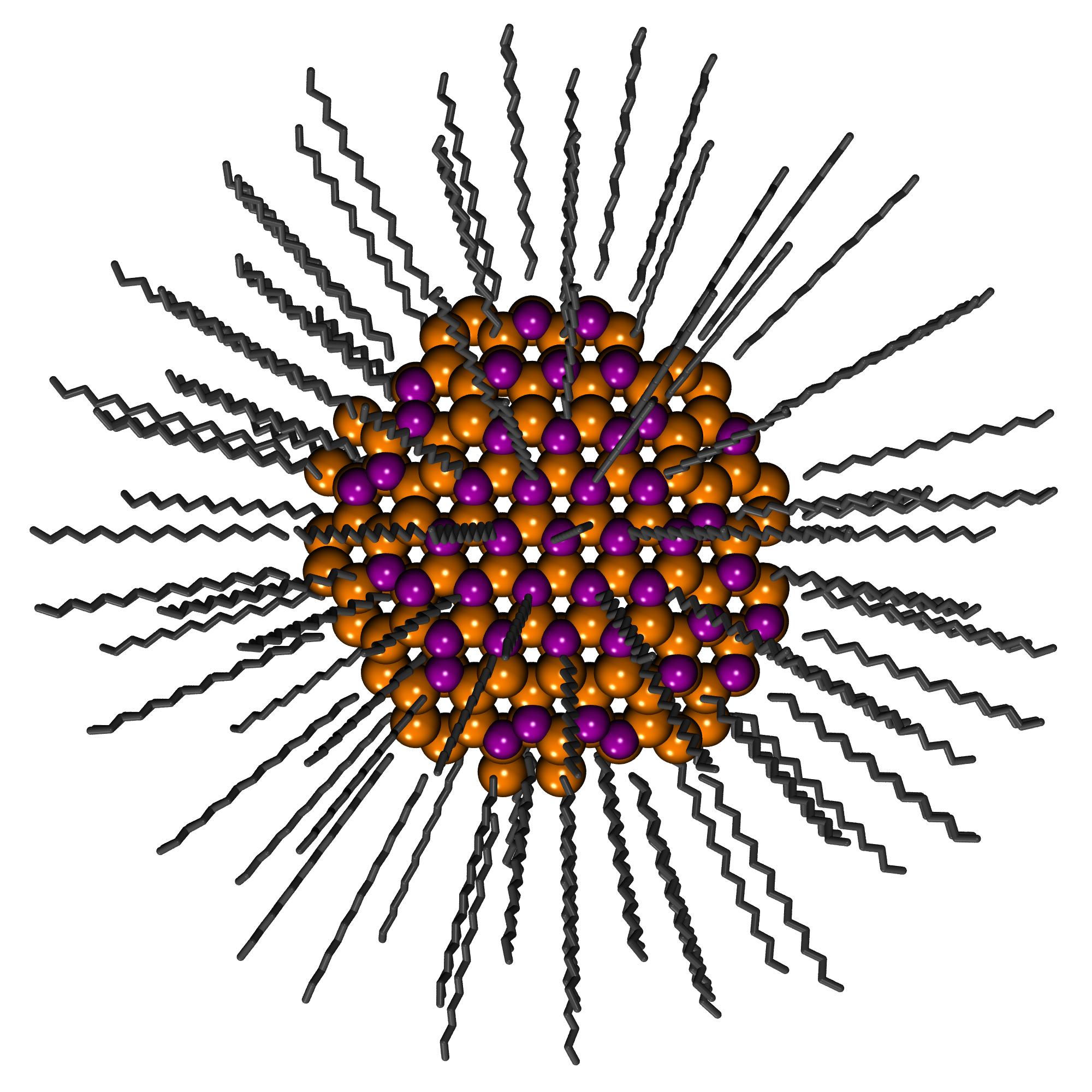}%
}%
}

\makebox[0.5\textwidth][l]{%
\subfigure[\mbox{}]{%
\includegraphics[width=0.25\textwidth]{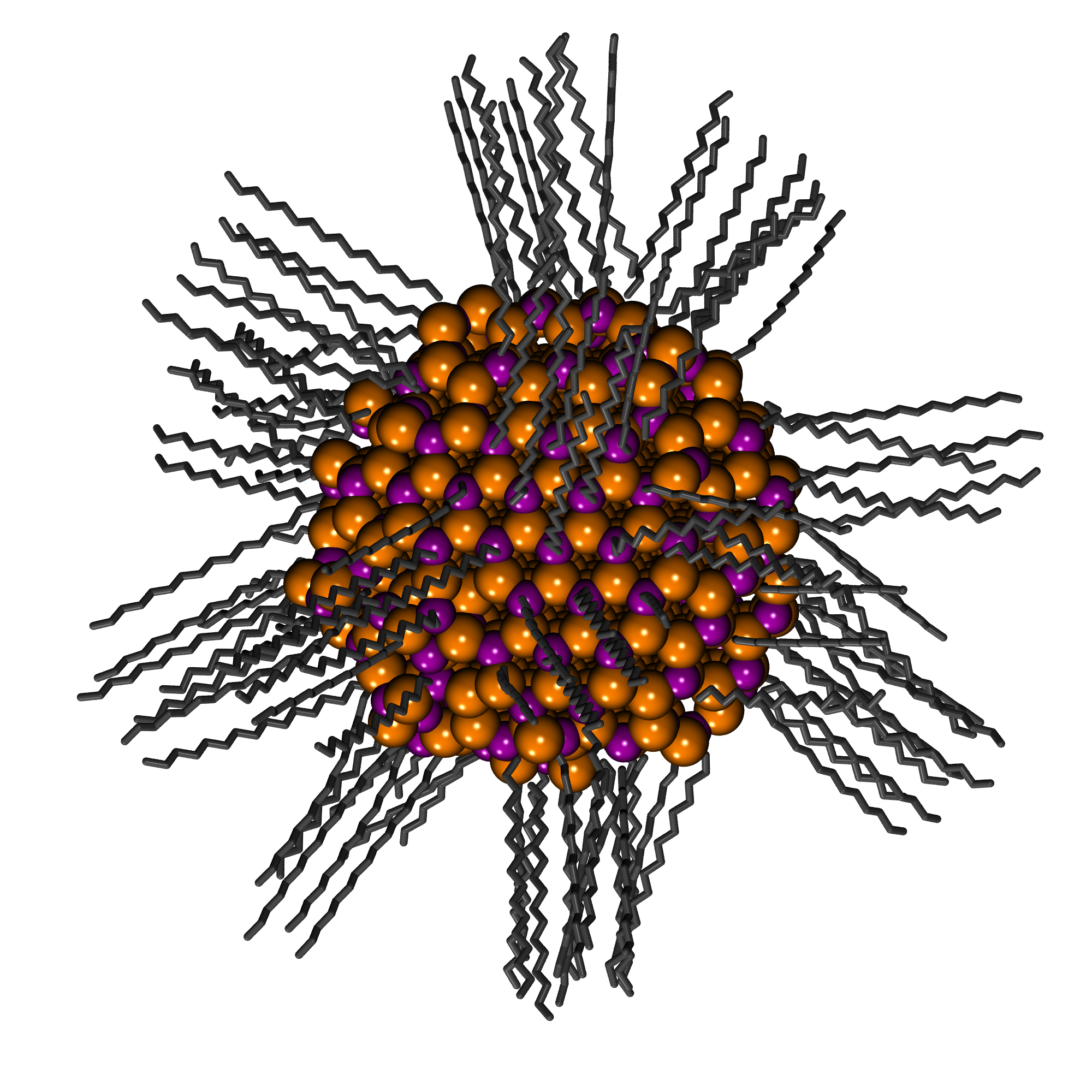}
}%
\subfigure[\mbox{}]{%
\includegraphics[width=0.25\textwidth]{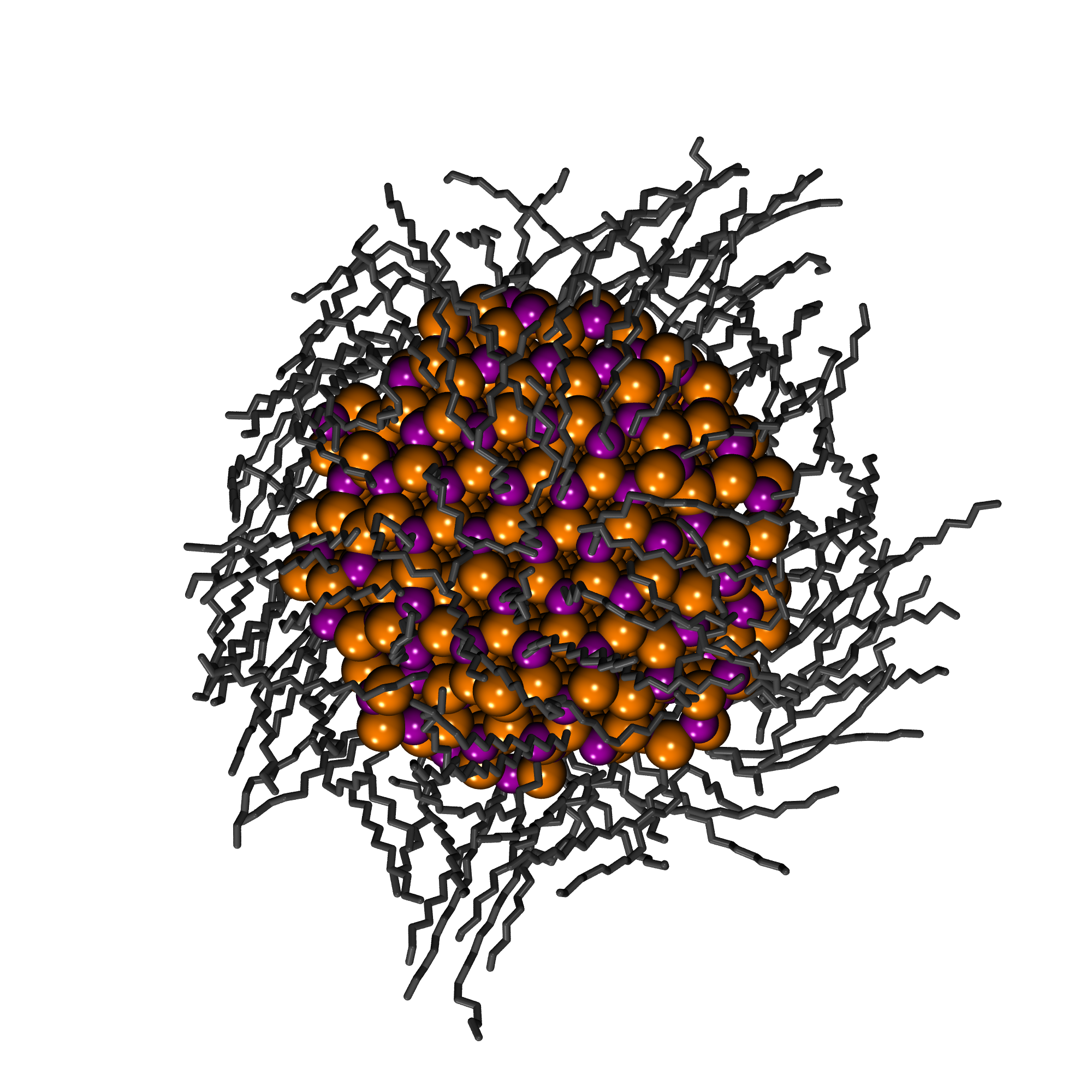}\label{1d}
}%
}

\caption{Steps of the simulation are ordered as follows:
a) the nanocrystal as carved from bulk; 
b) which is then decorated with ligands in a `spiky ball' conformation;
c) the geometry after energy minimization;
d) the geometry after molecular dynamics}
\label{fig:methods-picture}
\end{figure}

Figure \ref{fig:methods-picture} shows the simulation protocol we followed. 
The nanocrystal core structure is carved from the bulk Wurtzite crystal structure. 
We remove surface atoms with only a single bond while maintaining the stoichiometry. 
The nanocrystal is then decorated in a configuration in which every ligand points
directly away from the center of the nanocrystal, with the head of the ligand
placed a Cd-N bond distance away from a surface Cd. 
This initial structure owes its inspiration to the conventional picture for the ligand structure around a nanocrystal, the so-called `spiky ball'~\cite{weidman_interparticle_2015}. We then perform a minimization
step to reduce stress from unfavorable surface conditions and nearby
ligands. This configuration then undergoes a molecular dynamics simulation to get
the structure of the ligands at room temperature. Finally, we remove any ligand whose head group is more than 0.3 nm away from the surface, which we deemed to be detached from the surface during the simulation.
In this manner, coverage of 85-95\% available surface sites was achieved, which is comparable with experiment~\cite{anderson_ligand_2013}.

We simulated the same nanocrystal with ligands of varying lengths, from 0.30\,nm to
2.5\,nm, since a key engineering aspect of organic shell ligands is the ability to change the length of the ligands through ligand exchange reactions~\cite{anderson_ligand_2013}.
We also chose three initial nanocrystal cores of sizes 1.00\,nm, 1.78\,nm, and 2.50\,nm in radius to study the influence of dot size on the conclusions. 
One core structure per size was carved from bulk CdSe to yield equal stoichiometry and used for all simulations for this dot size. 
In this way, we focus on the effect of the ligands themselves without being concerned with variability in the core structure.

\begin{figure}
\includegraphics{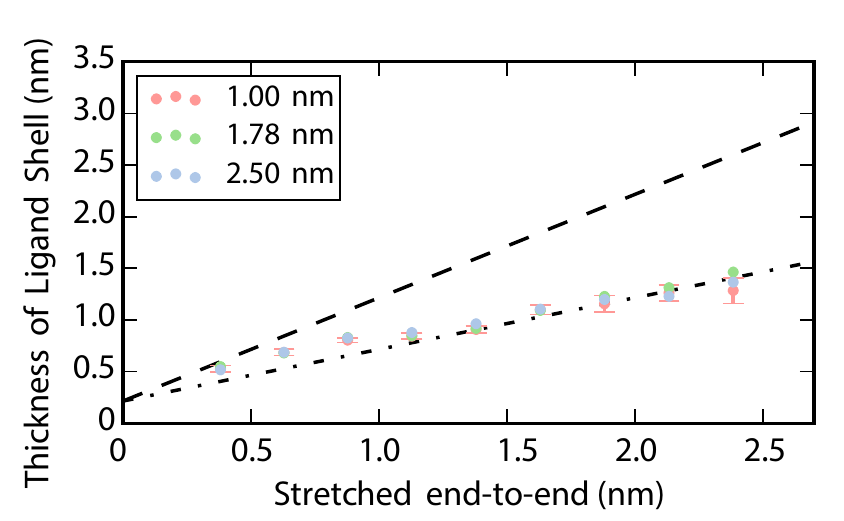}
\caption{Thickness of the ligand shell as a function of stretched ligand length for three dot sizes. 
Two linear eye-guides are also provided: the first (-~-~-) has a
slope of one, shifted by the Cd-N bond length: $x+b_{\text{Cd-N}}$.
The second (-.-.) has a slope one half: $x/2+b_{Cd-N}$. 
Representative error bars are shown for the smallest dot size, where we averaged over six independent runs of \reffig{fig:methods-picture}.
These show that the shell thickness does not vary significantly with radius over the range shown here. 
}
\label{fig:thickness-picture}
\end{figure}

\begin{figure}
\begin{center}

\makebox[0.5\textwidth][l]{%
\subfigure[\mbox{}]{%
\includegraphics[width=0.5\textwidth]{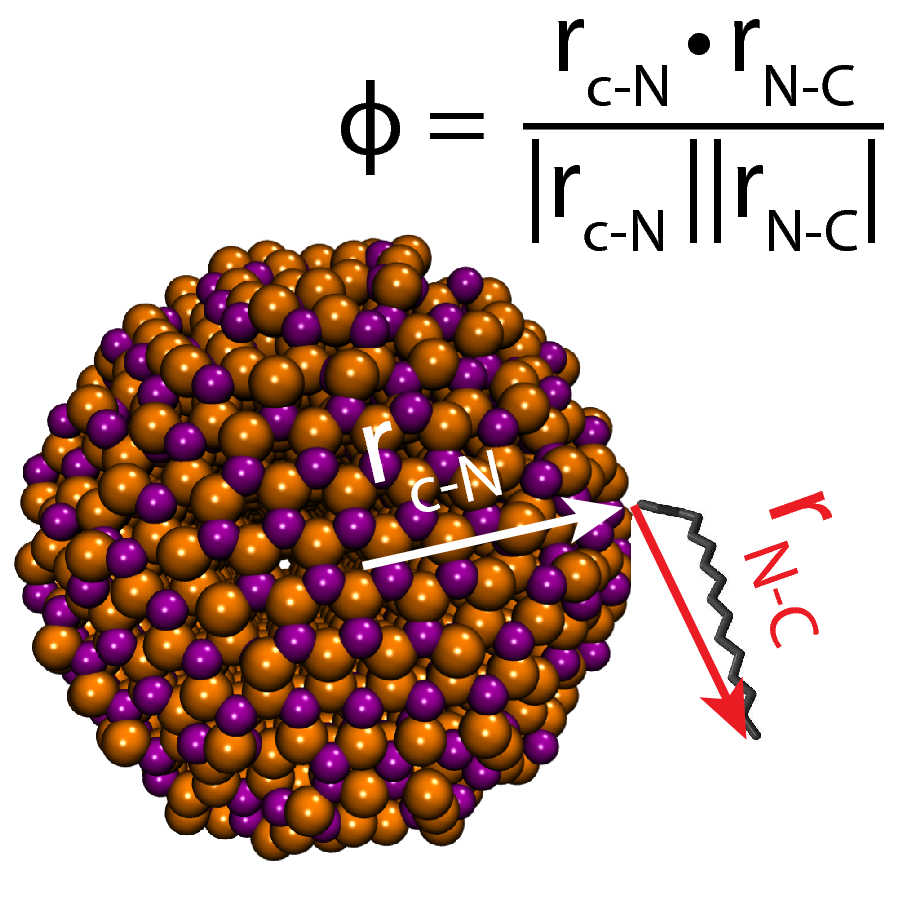}
\label{3a}
}
}

\makebox[0.5\textwidth][l]{%
\subfigure[\mbox{}]{%
\includegraphics[width=0.5\textwidth]{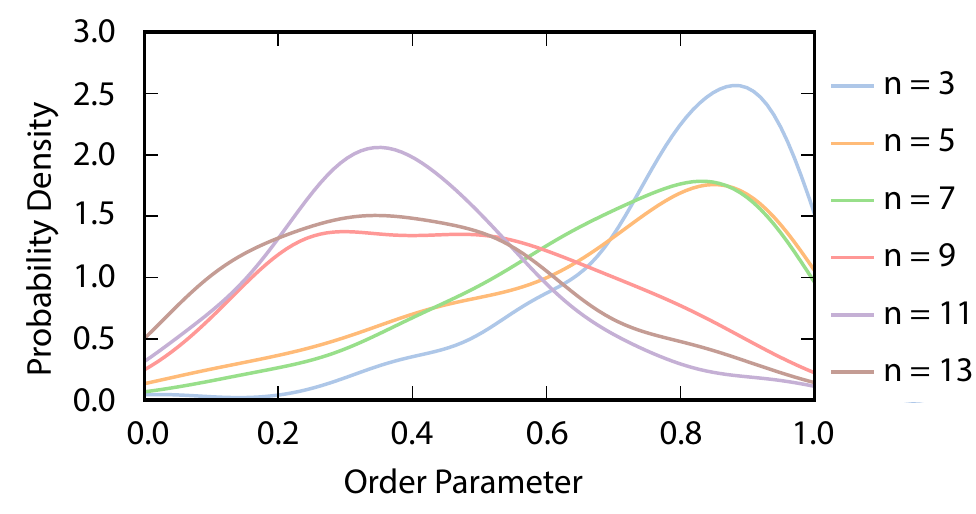}
\label{3b}
}
}

\caption{Calculation of the order parameter: For each ligand, we calculate two vectors: the first is the vector from the geometric
center of the nanocrystal to the ligand head group, the second is
the vector from the head group to the tail group (shown in (a)). The order parameter
is then the dot product of the normalized vectors. This distribution varies according to ligand length (shown in (b)), with ligand repeat units shown in the key as $n$ in CH$_3$(CH$_2$)$_n$NH$_2$. The dot size is 1.78\,nm.}
\label{fig:The-distribution-of}

\end{center}
\end{figure}

\begin{figure*}[t]

\includegraphics[width=1\textwidth]{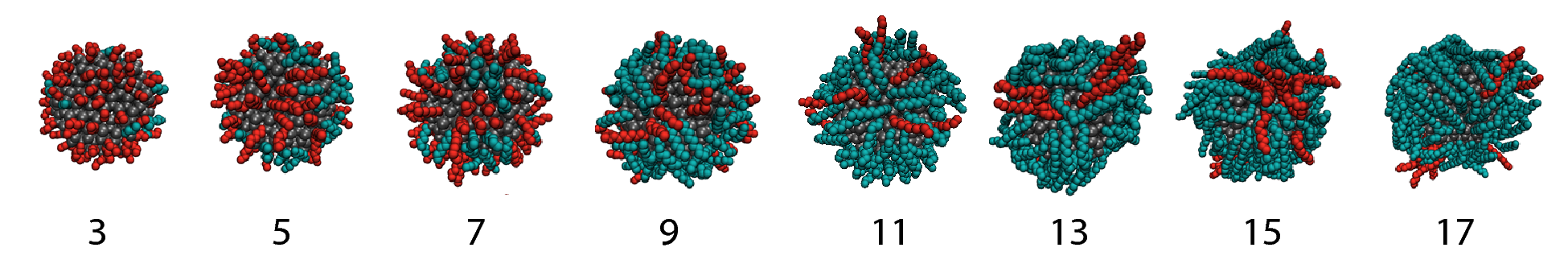}

\caption{Illustrative snapshots from simulations of a 1.78nm nanocrystal. 
From left to right, these correspond to ligand lengths of n=3 to 17 in increments of two.
The surface ligands are colored by
their orientation relative to the surface. Red-colored molecules are sticking straight out, as quantified by an order parameter greater than 0.6,
whereas turquoise-colored molecules lay flat. }
\label{fig:Illustrative-snapshots-from}
\end{figure*}

\begin{figure}[]
\includegraphics{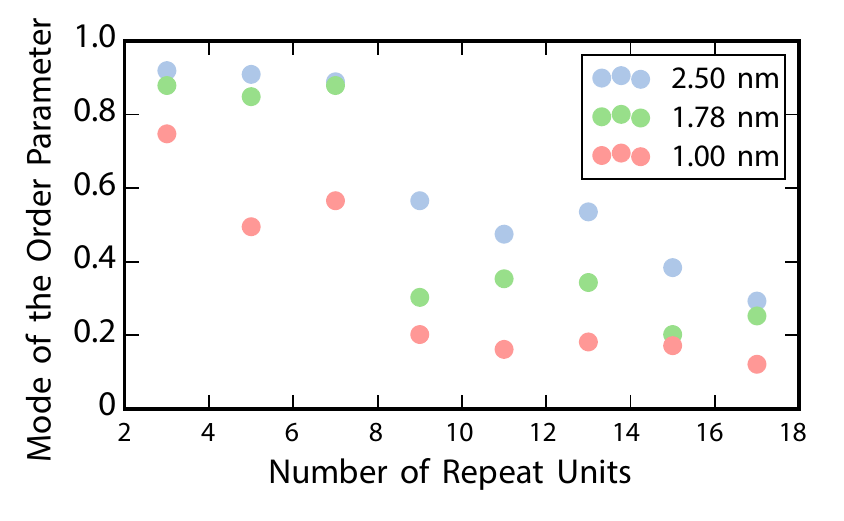}
\caption{Mode of the order parameter distribution as a function of ligand size,
for three nanocrystal sizes.\label{fig:3dot-order}}
\end{figure}

\begin{figure}
\includegraphics[width=1\textwidth]{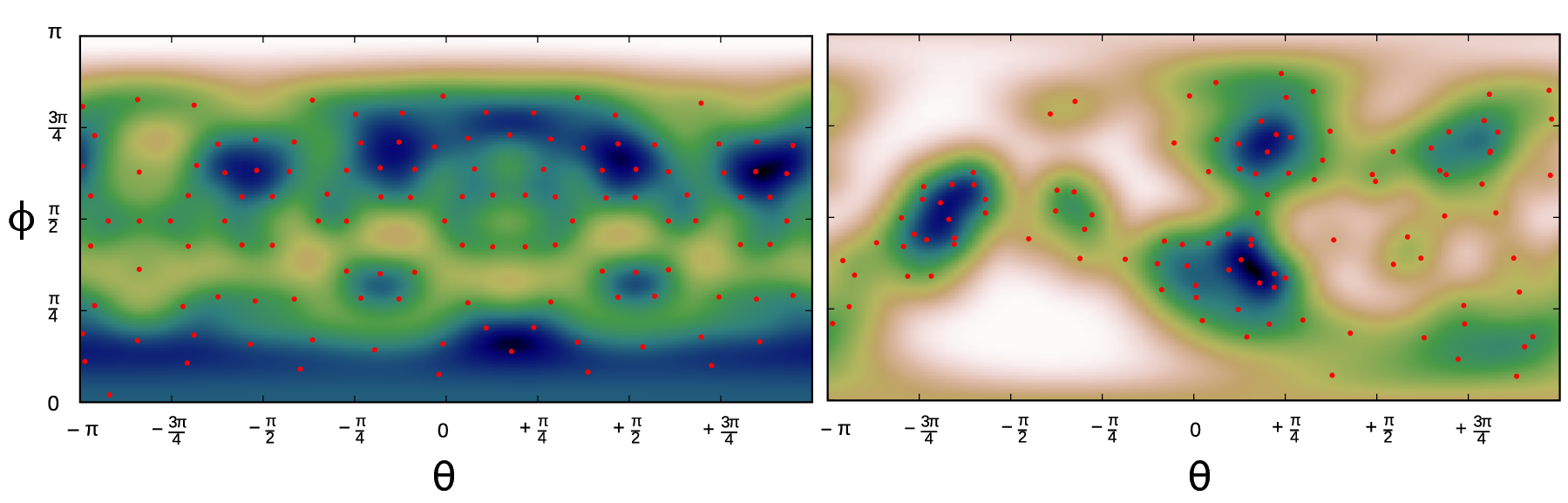}
\caption{At the start of the simulation (left) the height distribution is uniform within projection error but over the course of the simulation it becomes increasingly anisotropic (right). 
Red points indicate location of the end group (C in CH$_3$), which is where the height is measured. 
This measure is made continuous (see text for details). 
Here, blue locations are increased ligand height, white are decreased. The dot size is 1.78\,nm. }
\label{fig:heat-map-of}
\end{figure}

A simple measurement to take for our nanocrystals following molecular dynamics is the thickness of the ligand shell.
The thickness distribution for one dot is measured as the distance between the tail group (C in CH$_3$) and the closest Cd or Se atom. The effective thickness is then calculated as the average of these measurements. 
We find that shell thickness increases with ligand length as expected, but there is clearly a sub-linear component  large ligand lengths.
When plotted against ligand length, \reffig{fig:thickness-picture} shows that the thickness is approximately piecwise linear.
The slope changes from approximately unity to approximately one half between 0.5 nm~to 10 nm.

One can imagine several possibilities to explain this transition: perhaps the ligands are doubled over on themselves,  or lay flat against the surface, or form tight coils. 
By inspection (e.g. in \reffig{1d}), the ligands tend to lay flat. 
To quantitatively analyze this behavior, we constructed an order parameter
for the ligand arrangement. 
For each ligand, we calculate two vectors: the first is the vector from the geometric
center of the nanocrystal to the ligand head group, the second is
the vector from the head group to the tail group. The order parameter
is then the dot product of the normalized vectors. 
Figure \ref{3a}
illustrates the calculation of the order parameter. 
This construction approximates
the cosine of the angle between the ligand and the normal to the surface of the nanocrystal and was chosen
to be simple and easy to calculate, while still able to capture the
current state of the system. When a ligand sticks straight out of and is normal to the surface, its order parameter is 1,
and when the ligand lies completely flat and it is tangential to the
surface its order parameter is 0.

Figure \ref{3b} shows the distribution of this order parameter for a representative sample of ligand lengths in this study.
At short ligand lengths (3-7 repeat units) the ligands have a high order parameter of around 0.9. 
As the ligands grow longer (9-13 repeat units) a transition occurs, and most ligands have a low order parameter of 0.3 because they lie flat. 

The order parameter forms a useful tool for the visual inspection of the structures that we found. 
Rendered images of the dots are shown in \reffig{fig:Illustrative-snapshots-from}, where we have colored the ligands on the surface of the dot by their order parameter. 
Figure \ref{3b} allows us to choose a sensible dividing value for the change in color; red ligands have an order parameter greater than 0.6, while turquoise ligands have an order parameter less than 0.6.

We can therefore relate the observed conformation change to the surface thickness relationship found in \reffig{fig:thickness-picture}. 
At a ligand length of around 1 nm, corresponding to chains with $n=7$ to $n=9$, the slope starts to transition in \reffig{fig:thickness-picture}, and this is then due to a change in ligand shell conformations shown in \reffig{fig:Illustrative-snapshots-from}. 
The leftmost nanocrystal, with the smallest ligand length, looks like a `spiky ball' with the ligands projecting upward from the surface of the dot. 
The rightmost nanocrystal, with the longest ligand length, looks like a `hair ball', with the ligands bunched together and lying flat on the surface. 
As we move from left to right, we can see a tendency for the ligands to be flatter against the surface (change from red to turquoise). 
When seen through the lens of this conformation change, the simple heuristic that the shell thickness is equal to the half the ligand length can be seen to be serendipitous and part of a larger transition.
When the ligands are much longer than the radius of the CdSe core, we would expect volume-filling effects to dominate and the linear relationship to break down asymptotically. 
But for physically realistic dot sizes and ligand lengths, the intermediate transition regime dominates.

When the dot size is changed, the same qualitative observations described above persist.
However, there is a quantitative difference in order parameter distribution. 
Figure \ref{fig:3dot-order} shows the effects of dot size on the transition between the spiky ball to the wet hair. 
As the dot grows bigger, we observe that for intermediate ligand lengths ($n=5-12$) there is a pronounced increase in the percentage of ligands either sticking straight out ($n=3-7$) or a more significant mix of straight out and lying flat ($n=9-11$) as compared to the smallest dot size.
This fits our intuition well, as, in the limit of a flat interface, we expect the ligands to stand completely upright, as they do in self assembled mono-layers~\cite{halperin_tethered_1992}.

Next, we investigate the anisotropy of the height distribution on the surface of the nanocrystal because anisotropy in thickness can affect the way in which nanocrystals assemble.
For example, if the dots were amorphous and roughly spherical, they would adopt a body centered cubic structure.
Figure \ref{fig:heat-map-of} shows a
height/heat map of the ligands at the start and end of the simulation.
This distribution is made continuous using the von Mises--Fisher  function~\cite{fisher_statistical_1987}. 
Beginning with the coordinate of each tail group (corresponding to the red points in \reffig{fig:heat-map-of}, and the C in the CH$_3$ group), ${\bf \bar{r} }_i$, a spherical gaussian of width $\kappa$ is centered at that point.
The continuous distribution is then the sum over the height-weighted functions as follows:
\begin{equation}
f_p  \left( {\bf r} ; {\bf \bar{r} } ,\kappa \right) = \frac{1}{N_\text{points}} \sum_i^{N_\mathrm{points}} h_i \exp  \left(  \kappa \, {\bf \bar{r} }_i \cdot { \bf r } \right)
\end{equation}
where $h_i$ are the heights, and ${\bf r}$ are points on the sphere.
Finally, the function on the surface of a sphere is projected onto a two-dimensional heat map shown in \reffig{fig:heat-map-of} using a Mercator projection. 

From \reffig{fig:heat-map-of} we can make the observation that the spatial distribution of the ligands around the dots is not isotropic. 
Initially, the ligands are distributed almost uniformly around the
nanocrystal. After the simulation, there is a clustering of the ligands, leaving
parts of the dots more exposed, while other parts become more crowded.
This suggests that both the length of the ligand and the average distance
to the surfaces are not fully adequate measures of the thickness of the
ligand shell, since even in a well-covered nanocrystal with a thick shell there exist patches of substantially lower thickness. 
This can be thought of in terms of anisotropy classes, with our nanocrystals making a transition in surface coverage (corresponding to class A anisotropy, as described by Glotzer et. al.\cite{glotzer_anisotropy_2007}).

\section*{Experiment}

\begin{figure}

\makebox[0.5\textwidth][l]{%
\subfigure[\mbox{}]{%
\includegraphics[width=0.5\textwidth]{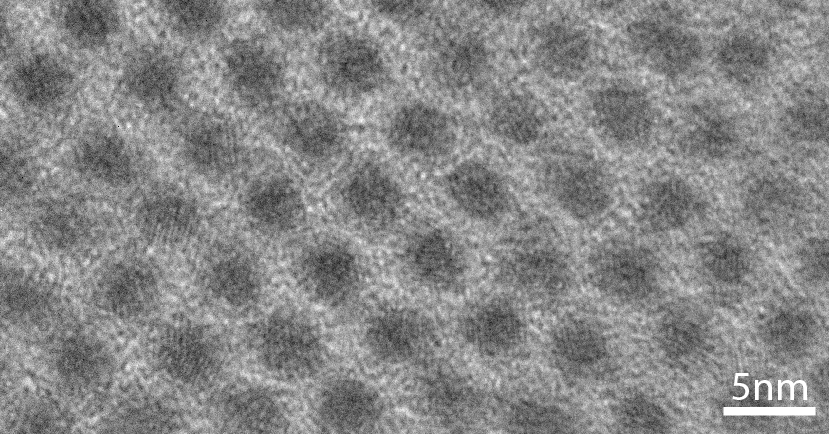}
\label{micrograph}
}
}

\subfigure[\mbox{}]{%
\includegraphics[width=0.5\textwidth]{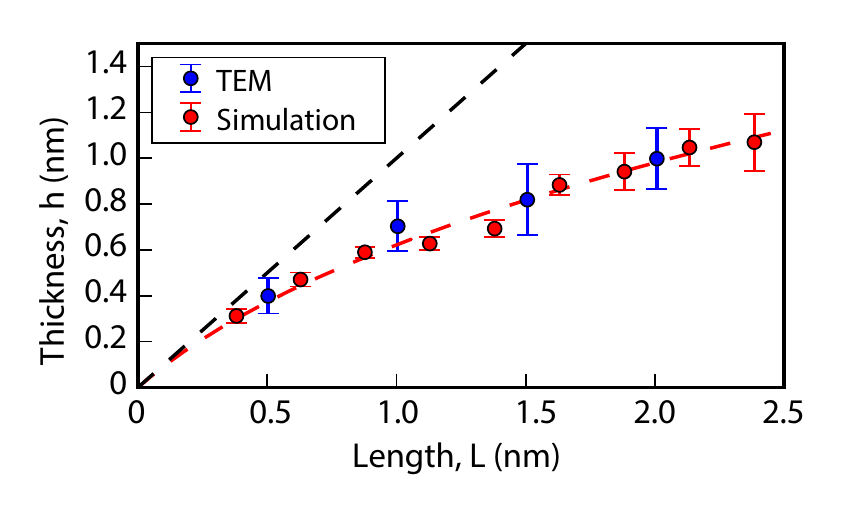}
\label{comparison}
}
\caption{Transmission electron microscopy (TEM) was performed to make comparison with computational results. (a) An example of a TEM micrograph of a layer of PbS nanocrystals with carboxylic acid ligands. (b) TEM data compares well with simulation data (here, taken from the 1.00\, nm dot size). The thickness in this plot is corrected for the bond length that joins the ligand to the surface \emph{i.e.} Cd-N or Pb-O. The red, dashed line shows the fit to \refeq{kappa} for the whole data set. The black, dashed line is a line of slope one.
}

\end{figure}

If the simulations are correct, the structure of the ligand sphere around a dot should have fairly clear implications in terms of how close a molecule or surface can get to a dot and how close two dots can get to each other. In this section, we test these implications by carefully examining the dot-to-dot spacing in a quantum dot array to determine if there is evidence for the spiky ball-to-hairball transition.

We synthesized lead sulfide quantum dots using a modified hot-injection method.~\cite{hendricks_tunable_2015}. In order to modify the ligands covering the surface, ligand exchange was performed ex-situ in toluene. We obtained the quantum dot arrays by drop-casting onto a TEM grid.
We analyzed TEM micrographs (\emph{e.g.} \reffig{micrograph}) by sampling the image intensity using a Fourier transform technique. The dot radius (d=2.67 +/- 0.35 nm) was subtracted from the peak-to-peak distance to yield the spacing between the edges of two dots, which amounts to double the ligand shell thickness.
To make comparison between the PbS dots synthesized and the CdSe dots simulated, the relevant ligand-surface bond distance was subtracted from the measurements -- Pb-O (0.23 nm)\cite{gourlaouen_[pbh2o]2+_2006} and Cd-N (0.215 nm, measured from the simulation) respectively.
The TEM micrographs and raw data can be found in the supplementary information.

We found the measurements of ligand shell thickness (\reffig{comparison}) agreed well with the simulations described in the previous section.
In particular, this applies both not only in the region between 1.2 nm and 2.0 nm, where the shell thickness is proportional to one-half the ligand length, but also at shorter lengths where transitionary behavior is seen.
We can thus conclude that it is very likely that the experimental quantum dots are undergoing the transition seen in our molecular dynamics simulations.

\begin{figure}

\includegraphics[width=0.45\textwidth]{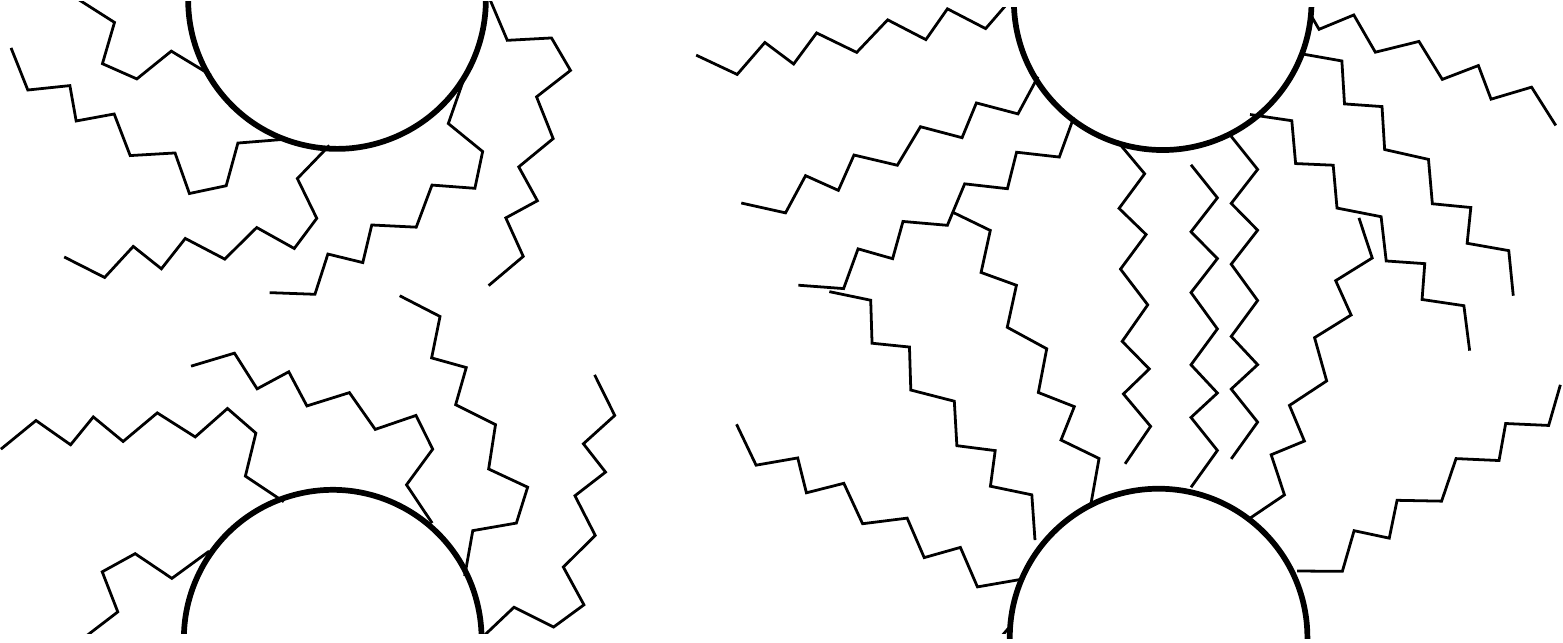}

\caption{Two explanations for how dots are separated by a distance of approximately one ligand length. Left panel shows interdigitation of the ligands, which is the prevailing literature viewpoint; right panel shows ligands crumpled over as in our simulations.}
\label{inter_vs_hairy}
\end{figure}

In previous experimental studies of nanocrystal-to-nanocrystal distance, the ligands are viewed to be interdigitated `spiky balls'.
This model was invoked in the explanation of X-ray experiments, which suggested that nanocrystals form superstructures with dot-to-dot distances comparable to a single stretched ligand length~\cite{weidman_interparticle_2015}.
In contrast, our `hairball' picture shows a different path to achieve this dot-to-dot distance, as two halves of a ligand plus two Cd-N bond lengths (see \reffig{inter_vs_hairy}).

There is also an important additional benefit to come from the experimental result.
The simulated system is CdSe dots with amine ligands, whilst the experimental data come from PbS dots with carboxylic acid ligands.
The agreement of ligand shell thickness beyond this disparity implies some degree of transferability of our findings to other systems which use alkyl ligands in a way that is insensitive to head group and semiconducting core.
To this end, we provide  the following heuristic to measure the distance, $d$, between two dots:
\begin{equation}
\begin{split}
d &= D_\mathrm{dot} + 2 b_\mathrm{dot-ligand} + 2 k \left( \left( 3 L/k+1 \right)^{\frac{1 }{3 }}-1\right); \\
k &\sim1.2 nm.
\label{kappa}
\end{split}
\end{equation}
In this equation $D_\mathrm{dot}$ is the diameter of the dot; $b_\mathrm{dot-ligand}$ is the bridging bond between the dot and the ligand. 
The final term represents twice the height of the ligand shell with ligands of length $L$, with $k$ a fitting parameter.
This functional form was chosen to reproduce limits of $h = L$ in the small $L$ limit and $h \propto L^{1/3}$ in the large $L$ limit, the latter corresponding to volume filling.
The parameter $k$ was found from fitting the data shown in \reffig{comparison}, and, overall, affords a more accurate way to calculate approximate dot-to-dot distances. While $k$ does depend on the size of the dot in principle, within the common size ranges, it seems reasonable to assume it is roughly constant.

\section*{Theory}

\begin{figure}
\begin{center}
\vspace{-0.5cm}
\makebox[1.0\textwidth][l]{%
\subfigure[\mbox{}]{%
\includegraphics{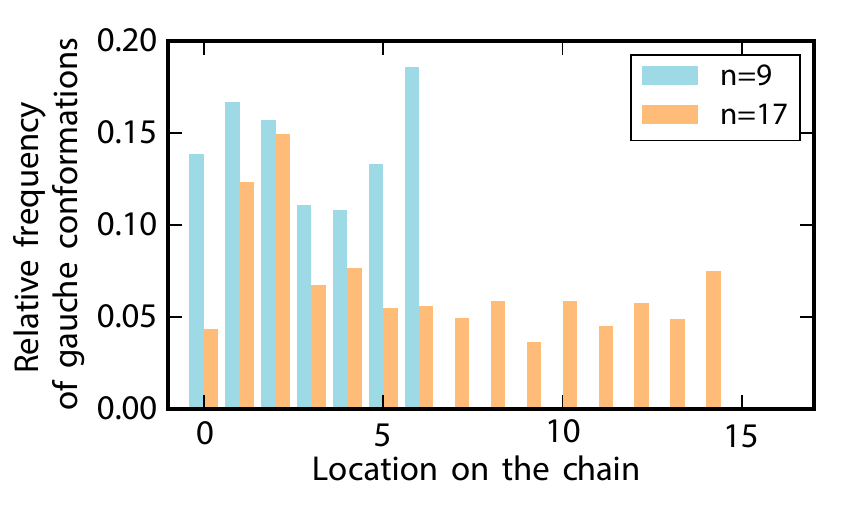}\label{fig:kinks}
}%
\subfigure[\mbox{}]{%
\includegraphics{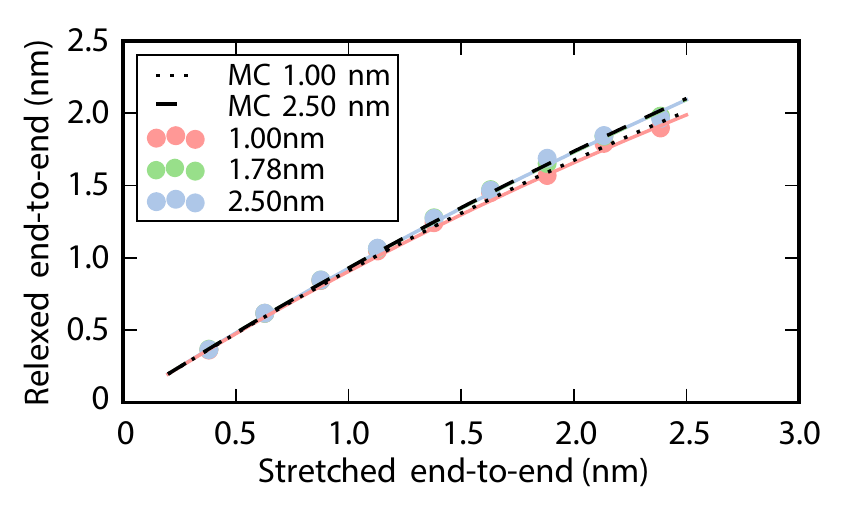}\label{fig:end-to-end-distance-vs.}
}
}
\subfigure[\mbox{}]{%
\includegraphics{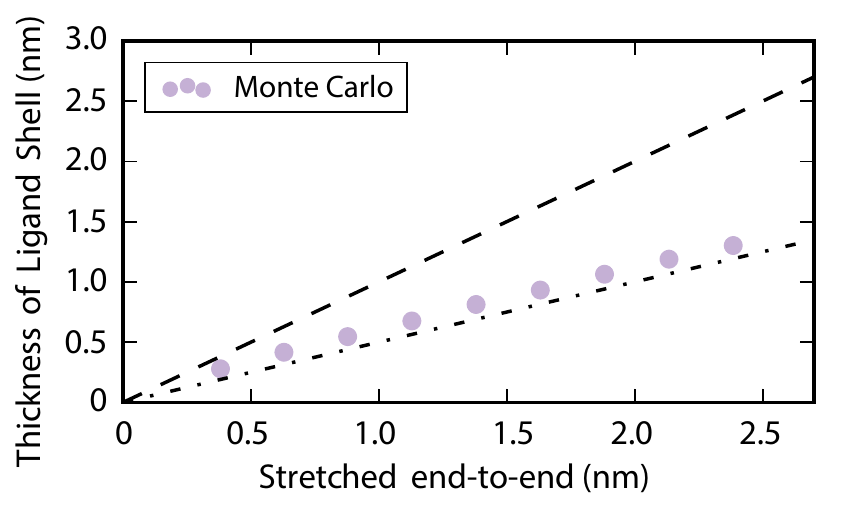}\label{fig:MCheights}
}
\caption{In (a), relative frequency of gauche conformations along the chain is plotted against chain location. 
Dihedrals are measured along the length of the chain, with chain location being referenced with the N-C-C-C dihedral as zero. 
For longer ligands, represented here by $n=17$, gauche conformations are more common at the beginning of the chain.
In (b), relaxed ligand end-to-end distance does not increase linearly with the stretched chain length of the ligand. 
Three dot sizes are shown. The red and green points overlay each other due to the similarity between the 1.78\,nm and 2.50\,nm dot.
The black, dashed lined labeled Monte Carlo (MC) provides a link between the dihedral angle and the end-to-end distance.
In (c), thickness of an imagined ligand shell is measured for the Monte Carlo simulation of a single ligand to show the contribution that is made to the thickness by the dihedral angles. A curve similar to the MD simulation is seen, despite the simplicity of the single-ligand picture which does not include volume-filling effects. The dashed line is a slope one linear scaling: $x$.
The dashed-dot line is a slope one half linear scaling: $x/2$.
}
\end{center}
\end{figure}

Given that simulations and experiments both predict similar behavior for the ligand sphere thickness, there might be some more fundamental theoretical explanation for what is occurring. 
Here, we theoretically analyze the ligands from a microscopic perspective. 
We sought to isolate the role of the dihedral angle, since the degree of flexibility of a polymeric chain is usually attributed to the freedom in the dihedral angle~\cite{flory_statistical_1969}.
There are minima in the potential at dihedral angles corresponding to $\pm$60 degrees (gauche) and 180 degrees (trans), with the gauche conformation being somewhat higher in energy.
Interatomic interactions mean that it can be favorable to form the gauche conformation, and this can cause substantial directional changes in the chain.

Figure \ref{fig:kinks} shows the distribution of gauche conformations along the chain for two representative dots.
The distribution of gauche conformations in the spiky ball is relatively uniform. 
In contrast, the `wet hair' structure has slight stabilization of early gauche conformations - that is to say, the chains twist preferentially near their base rather than in the middle or the end of the chain.
A similar result has been experimentally seen in a gold NC system\cite{badia_structure_1997}.
Steric crowding is unlikely to be an explanation for this trend, since it would favor dihedral changes further away from the dot. 
A better explanation is based on interchain attraction:
when one ligand lays flat, nearby ligands also have a tendency to lay flat to benefit from energetically favorable interchain interactions.
This would promote early distortion of the carbon backbone, and is our most plausible explanation for the distribution seen.
The ligand lies flat, rather than curling up, due to the presence of other neighbors with favorable intermolecular interactions.
Overall, then, this looks like `wet hair'.

If the dihedral angle distribution is indeed responsible for the surface thickness trend with increasing ligand length, 
then we should be able to devise a model which reproduces simulation observables with dihedral angles alone.
Towards this end, we make a Monte Carlo simulation of an isolated ligand based on rigid C--C bonds with constant bond angles, randomly sampling the dihedral angles in \reffig{fig:kinks}. 
Data shown in \reffig{fig:end-to-end-distance-vs.} demonstrates that the end-to-end distance of the ligands modeled in this way agrees with corresponding lengths collected from the molecular dynamics simulations.
When we use this dihedral angle distribution to measure the surface thickness due to the ligands (\reffig{fig:MCheights}), we find that the thickness values are consistent with the trend seen elsewhere in the manuscript.
This model involves only simulating the dihedral angles from a single ligand, they account for the \emph{average} effect of interactions with other ligands through the dihedral angle distributions (which were drawn from a model in which the ligands interact with one another).

To test whether the non-uniformity of the dihedral angles seen in \reffig{fig:kinks} is important for reducing surface thicknesses, dihedral angles are now sampled over both a uniform and a non-uniform distribution with the same average probability as shown in \reffig{fig:kinks}. 
For the 2.50\,nm nanocrystal with ligand length $n=17$ the average gauche conformation frequency is 12\%. 
Both the uniform and non-uniform distributions reproduce the results shown in \reffig{fig:end-to-end-distance-vs.}.
Therefore, while the average kink probability from the simulation is needed to reproduce the trend, a non-uniform distribution is not required.
This is unexpected - it suggests that perhaps weakly interacting ligands (or ligands that interact strongly with solvent) might also behave as wet hair, and this hypothesis is worth further testing.

\FloatBarrier

\section*{Conclusion}

In conclusion, we investigated the physical chemistry of the surface of a quantum dot with an emphasis on ligands. 
We performed molecular dynamics simulations on CdSe dots, treating the ligands and the core atomistically. 
We find that the ligands form a shell of approximately half the ligand length in thickness (on average) due to a transition that occurs between lengths of 0.5 nm and 1.5 nm. 
This was verified by transmission electron microscopy of PbS nanoarrays that determined a dot-to-dot separation consistent with the transition predicted by the simulations.
They are, instead, generally in an intermediate regime caused by flexibility in the dihedral angle and adoption of the gauche conformation. 
This causes the ligands to fall over on the surface of the dot, and we propose there is a `wet hair' appearance to the surface of the nanocrystal.

The average distance between a nanocrystal and a nearby object (e.g. a nearby 2D material surface, organic semiconductor, etc.) will be roughly half the length of the ligands for the range of typical ligand lengths; this drops off at longer ligand lengths as sub-linear behavior is seen.
These scaling relations suggest that transfer rates between dots and nearby objects will be enhanced due to the shorter distance, consistent with previous findings~\cite{thompson_energy_2014}.
We should also emphasize that because of the non-uniformity of the thickness of the ligand sphere, the distance of closest approach could be even smaller than the average distance of approach we have focused on here.
Such effects would likely have a modest effect on dot-to-dot energy or charge transfer (where bald spots would always be compensated for by thicker spots elsewhere) but could have a significant effect on molecule-to-dot transfer, where a single molecule on a bald spot could have greatly accelerated transfer.
Agreement in this paper between simulated CdSe/amine dots and experimental PbS/carboxylic acid dots shows that our findings are broadly applicable, and may well be general for ligands with alkyl groups. 

The availability of an accurate sense of where atoms are on the surface of a nanocrystal has advantages that go beyond the scope of this paper, and provide direction for further work. 
The role of the ligand shell in energy transfer to organic molecules and, ultimately, up-conversion is poorly understood; the atomic configurations we have developed through molecular dynamics will be of use in providing realistic interfaces for electronic structure calculations of couplings and transfer rates. 
The anisotropy that we see in the dots will also play an important role in how the dots pack and adopt their superstructure, which is important when considering which ligands to use in synthesis and the resultant quality of the arrays~\cite{weidman_interparticle_2015}. 

There are some obvious future directions suggested by the present study. The simulations contained only a single QD, and therefore do not include dot-to-dot interaction. 
There was also no solvent in the simulation, so the results for QDs in solution may be different. In particular, it would be interesting to see how solvent polarity influences the transition from a spiky ball to more of a wet hair configuration.
Finally, on the experimental side, with a quantitative understanding of the ligand sphere in hand it will be extremely interesting to return to the questions of fission and up-conversion and quantitatively assess the underlying rates as a function of ligand shell thickness. Such a study could give insight into the incorporation of molecules into the ligand shell structure.

\section*{Methods}

{\bf Molecular dynamics:--}
The  GROMACS software package was used to perform the MD calculations~\cite{berendsen_gromacs:_1995}.
For the organic ligands, the OPLS force field was used~\cite{jorgensen_development_1996}. 
The CdSe nanoparticle was modeled using the bulk phase force field parametrized by Rabani~\cite{rabani_interatomic_2002-1}. 
The time-step was 2~fs using a velocity Verlet integrator and an Anderson thermostat at 300 K in an NVT ensemble. 
Snapshots were taken at intervals of 40 ps resulting in 100 snapshots in total. 
Error bars were computed for the smallest dot size (1.00\, nm). 
Here, slightly different radii (0.98--1.02\, nm) were used to carve the dot giving rise to slightly different numbers of atoms in the $\text{Cd}_n\text{Se}_n$ core ( $n=66, 68, 69, 73, 75, 80$).

Following the methodology used by Schapotschnikow and coworkers~\cite{schapotschnikow_adsorption_2009}, we model the interaction relying on the Lennard-Jones parameters and the partial charges of the nanocrystal and ligand atoms \emph{i.e.} we chose not to include any explicit nanocrystal-ligand bonding terms.
While simple, Schapotschnikow demonstrated that this force field reproduces the experimental binding energy of the ligands to CdSe~\cite{schapotschnikow_adsorption_2009}. From our simulation, we also see that the force field preserves geometries such as the cadmium coordination with the nitrogen atoms.

A key part of the simulations is that the ligands and the nanocrystal atoms are all free to move. Since the nanocrystal atoms and the ligand head groups are only bound through pair-wise potentials, ligands can detach and the surface atoms can relax and re-arrange. 

All 3d images were generetad using VMD\cite{HUMP96} and Tachyon\cite{STON1998}.

{\bf PbS dot synthesis:--}
Lead sulfide quantum dots with a first excitonic peak at 790 nm are synthesized following a modified hot-injection method as reported elsewhere~\cite{hendricks_tunable_2015}. In particular, a three-neck roundbottom flask is charged with 2 g of $\text{Pb}(\text{oleate})_2$ in 20 ml octadecene (ODE) and degassed under vacuum for 12h at 120$^{\circ}$C, before it is backfilled with nitrogen. The temperature is then decreased to 90$^{\circ}$C and 0.27 ml trimethylsilylthiane in 4 mL ODE is quickly injected. The reaction is immediately quenched by an ice bath. The quantum dots are purified by a common solvent-nonsolvent procedure and stored as a concentrated solution in toluene in a nitrogen glovebox. Ligand exchange is performed by diluting 0.1 mL of the quantum dot stock solution to a total volume of 0.5 mL and adding 0.1 mL of a 0.1 M ligand stock solution in toluene (4C, 16C) or a 0.4 M ligand stock solution (8C, 12C). The reaction is stirred air-free at room temperature for 3h before purification via a standard solvent-nonsolvent procedure. 

{\bf Transmission electron microscopy:--}
Transmission electron microscopy (TEM) is performed on a JEOL 2010 microscope at varying magnifications. TEM samples are prepared by drop-casting the resulting ligand-exchanged quantum dot solution onto TEM grids (UC-A on holey 400 mesh Cu, Ted Pella).

\section{Acknowledgements}
This work was supported at part of the Center for Excitonics, an Energy Frontier Research Center funded by the U.S. Department of Energy, Office of Science, Basic Energy Sciences (BES) under award number: DE-SC0001088. 
This work made use of the MRSEC Shared Experimental Facilities at MIT, supported by the National Science Foundation under award number DMR-1419807.
One of us (JJS) would like to thank the Royal Commission for the Exhibition of 1851 (UK) for a Research Fellowship.
We would like to thank M. Wilson for discussions.

Supporting Information Available: Electron micrographs, analyses and tabulated experimental data. This material is available free of charge via the Internet at http://pubs.acs.org.

%\bibliography{top2,extra}
\end{document}